\documentclass[a4paper]{article}

\usepackage{INTERSPEECH2020}
\usepackage{xcolor}

\title{On the Comparison of Popular End-to-End Models for Large Scale Speech Recognition}
\name{Jinyu Li$^1$, Yu Wu$^2$, Yashesh Gaur$^1$, Chengyi Wang$^2$, Rui Zhao$^1$, Shujie Liu$^2$}
\address{
  $^1$Microsoft Speech and Language Group\\
  $^2$Microsoft Research Asia}
\email{\{jinyli, wu.yu, yagaur, v-chengw, ruzhao, shujiliu\}@microsoft.com}

\begin{document}

\maketitle
\begin{abstract}
Recently, there has been a strong push to transition from hybrid models to end-to-end (E2E) models for  automatic speech recognition. Currently, there are three promising E2E methods: recurrent neural network transducer (RNN-T), RNN attention-based encoder-decoder (AED), and Transformer-AED.
In this study, we conduct an empirical comparison of RNN-T, RNN-AED, and Transformer-AED models, in both non-streaming and streaming modes. We use 65 thousand hours of Microsoft  anonymized training data to train these models. As E2E models are more data hungry, it is better to compare their effectiveness with large amount of training data. To the best of our knowledge, no such comprehensive study has been conducted yet. We show that although AED models are stronger than RNN-T in the non-streaming mode, RNN-T is very competitive in streaming mode if its encoder can be properly initialized. Among all three E2E models, transformer-AED achieved the best accuracy in both streaming and non-streaming mode. We show that both streaming RNN-T and transformer-AED models can obtain better accuracy than a highly-optimized hybrid model.
\end{abstract}
\noindent\textbf{Index Terms}: end-to-end, RNN-transducer, attention-based encoder-decoder, transformer

\section{Introduction}
Recently, the speech community is seeing a significant trend of moving from deep neural network based hybrid modeling \cite{DNN4ASR-hinton2012} to end-to-end (E2E) modeling \cite{miao2015eesen, chan2016listen, prabhavalkar2017comparison, battenberg2017exploring,  rao2017exploring, chiu2018state, Li18CTCnoOOV, he2019streaming, Li2020Developing} for automatic speech recognition (ASR). While hybrid models require disjoint optimization of separate constituent models such as acoustic and language model, E2E ASR systems directly translate an input speech sequence into an output token (sub-words, or even words) sequence using a single network.   

Some widely used contemporary E2E approaches for sequence-to-sequence transduction are: (a) Connectionist Temporal Classification (CTC) \cite{graves2006connectionist, Graves-E2EASR}, (b) recurrent neural network Transducer (RNN-T)\cite{Graves-RNNSeqTransduction}, and (c) Attention-based Encoder-Decoder (AED) \cite{Attention-bahdanau2014, Attention-speech-chorowski2015, chan2016listen}. Among these three approaches, CTC was the earliest and can map the input speech signal to target labels without requiring any external alignments. However, it also suffers from the conditional frame-independence assumption.  RNN-T extends CTC modeling by changing the objective function and the model architecture to remove the frame-independence assumption.  
Because of its streaming nature, RNN-T has received a lot of attention for industrial applications and has also managed to replace traditional hybrid models for some cases \cite{he2019streaming, Sainath19, Li2019RNNT, jain2019rnn}. 

AED is a general family of models that was initially proposed for machine translation \cite{bahdanau2014neural} but has shown success in other domains (including ASR \cite{Attention-bahdanau2014, Attention-speech-chorowski2015, chan2016listen}) as well. These models are not streaming in nature by default but there are several studies towards that direction, such as monotonic chunkwise attention \cite{chiu2017monotonic} and triggered attention \cite{moritz2019triggered}. The early AED models used RNNs as a building block for its the encoder and decoder modules. We refer to them as RNN-AED in this study. More recently, the transformer architecture with self attention \cite{vaswani2017attention} has also become prevalent and is being used as a fundamental building block for encoder and decoder modules \cite{dong2018speech, zhou2018syllable, karita2019comparative}. We refer to such a model as Transformer-AED in this paper.  

Given the fast evolving landscape of E2E technology, it is timely to compare the most popular and promising E2E technologies for ASR in the field, shaping the future research direction. This paper focuses  on the comparison of  current most promising E2E technologies, namely RNN-T, RNN-AED, and Transformer-AED, in both non-streaming and streaming modes. All models are trained with 65 thousand hours of Microsoft  anonymized   training data. As E2E models are data hungry, it is better to compare its power with such a large amount of training data.  To our best knowledge, there is no such a detailed comparison. In a recent work \cite{Sainath19}, the streaming RNN-T model was compared with the non-streaming RNN-AED. In \cite{chiu2019comparison}, streaming RNN-AED is compared with streaming RNN-T for long-form speech recognition. In \cite{karita2019comparative}, RNN-AED and Transformer-AED are compared in a non-streaming mode, with training data up to 960 hours. As the industrial applications usually requires the ASR service in a streaming mode, we further put more efforts on how to develop these E2E models in a streaming mode. While it has been shown in \cite{sainath2020streaming} that combining RNN-T and RNN-AED in a two-pass decoding configuration can surpass an industry-grade state-of-the-art hybrid model,  this study shows that a single streaming E2E model, either RNN-T or Transformer-AED, can also surpass a state-of-the-art hybrid model \cite{li2020high, li2019improving}.  
 
In addition to performing a detailed comparison of these promising E2E models for the first time, other contributions of this paper are 1) We propose a multi-layer context modeling scheme to explore future context with significant gains; 2) The cross entropy (CE) initialization is shown to be much more effective than CTC initialization to boost RNN-T models; 3) For streaming Transformer-AED, we show chunk-based future context integration is more effective than the lookahead method; 4) We release our Transformer related code with reproducible results on Librispeech at \cite{Wang2020Transformer} to facilitate future research. 
 
\section{Popular End-to-End Models}
\label{sec:e2e}

In this section, we give a brief introduction of current popular E2E models: RNN-T, RNN-AED, and Transformer-AED. These models have an acoustic encoder that generates high level representation for speech and a decoder, which autoregressively generates output tokens in the linguistic domain. While the acoustic encoders can be same, the decoders of RNN-T and AED are different. In RNN-T, the generation of next label is only conditioned on the label outputs at previous steps while the decoder of AED conditions the next output on acoustics as well. More importantly, RNN-T works in a frame-synchronized way while AED works in a label-synchronized fashion.


\subsection{RNN transducer}
The encoder network converts the acoustic feature $x_{1:T}$ into a high-level representation $h_{1:T}^{enc}$.
The decoder, called prediction network, produces a high-level representation $h_u^{pre}$ by consuming previous non-blank target $y_{u-1}$. Here $u$ denotes output label index. 
The joint network is a feed-forward network that combines the encoder network output $h_t^{enc}$ and the prediction network output $h_u^{pre}$ to generate the joint matrix $h_{t,u}$, which is used to calculate softmax output. Here $t$ denotes time index. 

The encoder and prediction networks are usually realized using RNN with LSTM \cite{Hochreiter1997long} units. When the encoder is a unidirectional LSTM-RNN as  Eq. \eqref{eq:uni_enc}, RNN-T works in streaming mode by default.

\begin{equation}
h_t^{enc} = LSTM (x_t, h_{t-1}^{enc})	\label{eq:uni_enc}
\end{equation}
However, when the underlying LSTM-RNN encoder is a bi-directional model as Eq. \eqref{eq:bi_enc}, it is a non-streaming E2E model. 
\begin{equation}
h_t^{enc} = [LSTM (x_t, h_{t-1}^{enc}), LSTM (x_t, h_{t+1}^{enc})]	\label{eq:bi_enc}
\end{equation}
When implemented with LSTM-RNN, the prediction network formulation is 
\begin{equation}
h_u^{pre} = LSTM (y_{u-1}, h_{u-1}^{pre}).		\label{eq:dec}
\end{equation}

With the advantage of Transformer models, there is a recent work to replace the LSTM-RNN in the encoder with the Transformer model to construct Transformer transducer \cite{zhang2020transformer} and Conformer transducer \cite{gulati2020conformer}.

\subsection{Attention-based Encoder-Decoder}
While RNN-T has received more attention from the industry due to its streaming nature, the Attention-based Encoder-Decoder (AED) models attracts more research from academia because of its powerful attention structure. 
RNN-AED and Transformer-AED differ at the realization of encoder and decoder by using LSTM-RNN and Transformer, respectively.

\subsubsection{RNN-AED}
 The encoder of RNN-AED can have the same structure as RNN-T like Eq. \eqref{eq:uni_enc} and Eq. \eqref{eq:bi_enc}. However, the attention-enhanced decoder operates differently as below: 
\begin{equation}
h_u^{dec} = LSTM (c_u, y_{u-1}, h_{u-1}^{dec}).		\label{eq:rnnaed_dec}
\end{equation}
here $c_u$ is the context vector obtained by weighted combination of the encoder output. $c_u$ is supposed to contain the acoustic information necessary to emit the next token. It is calculated using the help of the attention mechanism \cite{Attention-bahdanau2014, bahdanau2016end}.





\subsubsection{Transformer-AED}
Even though RNNs can capture long term dependencies, Transformer \cite{vaswani2017attention} based models can do it more effectively given the attention mechanism sees all context directly. Specifically, the encoder is composed of a stack of Transformer blocks, where each block has a multi-head self-attention layer and a feed-forward layer. Suppose that the input of a Transformer block can be linearly transformed to $\mathbf{Q}$, $\mathbf{K}$, and $\mathbf{V}$. Then, the output of a multi-head self-attention layer is 
\begin{flalign}\label{eq:transformer}
\text{Multihead}(\mathbf{Q,K,V}) &= [\mathbf{H_1} \ldots \mathbf{H}_{d_{head}}]\mathbf{W}^{head}\\
\text{where~}  \mathbf{H_i} &= \text{softmax}(\frac{\mathbf{Q_iK_i^{T}}}{\sqrt{d_k}})\mathbf{V_i}, 
\nonumber
\\
\mathbf{Q_i} = \mathbf{Q} W^{Q_i}, \mathbf{K_i}& = \mathbf{K} W^{K_i}, \mathbf{V_i} = \mathbf{V} W^{V_i}.
\nonumber 
\end{flalign}
 Here  $d_{head}$ is the number of attention heads and $d_k$ is the dimension of the feature vector for each head. This output is fed to the feed-forward layer. Residual connections \cite{RESNET-he2015} and layer normalization (LN) \cite{ba2016layer} are indispensable when we connect different layers and blocks. 
In addition to the two layers in an encoder block, the  Transformer decoder also has an additional third layer that performs multi-head attention over the output of the encoder. This is similar to the attention mechanism in RNN-AED. 

\section{Our Models}

\subsection{Model building block}
\label{ssec:block}
The encoder and decoder of E2E models are constructed as the stack of multiple building blocks described in this section.  For the models using LSTM-RNN, we explore two structures. The first one, LSTM\_cuDNN, directly calls Nvidia cuDNN library \cite{chetlur2014cudnn} for the LSTM implementation. We build every block by concatenating a cuDNN LSTM layer, a linear projection layer to reduce model size, and then followed by LN. Calling Nvidia cuDNN implementation enables us for fast experiment of comparing different models.

The second structure, LSTM\_Custom, puts LN and projection layer inside LSTM, as it was indicated in \cite{he2019streaming} that they are important for better RNN-T model training. Hence, we only use this structure for RNN-T by customizing the LSTM function. The detailed formulations are in \cite{Li2019RNNT}. However, this slows down the model training speed by 50\%.

For the Transformer-AED models,  we remove the position embedding part \cite{wang2020semantic} and use a VGG-like convolution module \cite{simonyan2014very} to pre-process the speech feature $x_{1:T}$ before the Transformer blocks. The LN is put before multi-head attention layer (Pre-LN), which makes the gradients well-behaved at the early stage in training.

\subsection{Non-streaming models}
We achieve non-streaming behavior in RNN-T by adding bidirectionality in the encoder. The encoder of this RNN-T is composed of multiple blocks of bi-directional LSTM\_cuDNN as described in Section \ref{ssec:block}. The prediction network is realized with multiple uni-directional blocks of LSTM\_cuDNN.

Similar to RNN-T, the non-streaming RNN-AED investigated in this study also uses multiple blocks of bi-directional LSTM\_cuDNN in the encoder and  uni-directional  LSTM\_cuDNN in the decoder. This decoder works together with a location-aware softmax attention  \cite{Attention-speech-chorowski2015}. No multi-task training or joint-decoding with CTC is used for RNN-AED.

Following \cite{karita2019comparative}, the Transformer-AED model uses the multi-task training and  the joint decoding of CTC/attention. The training objective function is 
\begin{equation}
\mathcal{L} = - \alpha \log p_{ctc}(y|x_{1:T}) - (1-\alpha) \log p_{att}(y|x_{1:T}).
\end{equation} 
The log-likelihood of the next subword $\log p(y_u|x_{1:t},y_{1:u})$ in the joint decoding is formulated as  
\begin{equation} \label{eq:joint_decoding}
  \log p_{ctc}(y_u|x_{1:t},y_{1:u})  + \beta_1 \log p_{att}(y_u|x_{1:t},y_{1:u}).
\end{equation} In practice, we first use the attention model to select top-k candidates and then re-rank them with Eq. \ref{eq:joint_decoding}.

\subsection{Streaming models} \label{sec:streaming_model}

Streaming RNN-T model has a uni-directional encoder. While we can directly incorporate a standard LSTM as the building block with either LSTM\_cuDNN or LSTM\_Custom as described in Section \ref{ssec:block}, incorporating the future context into encoder structure can significantly improve the ASR accuracy, as shown in \cite{Li2019RNNT}. However, different from \cite{Li2019RNNT} which explores future context frames together with the layer trajectory structure, in this study we propose to only use context modeling. We do this to save model parameters. Future context is modelled using the simple equation below. 
\begin{equation}
{\zeta}_{t}^{l} = \sum_{\delta =0}^{\tau}  {q}_{\delta}^{l} \odot {g}_{t+\delta}^{l} \label{eq:context}. 
\end{equation}
Because $\odot$ is element-wise product, Eq. \eqref{eq:context} only increases the number of model parameters very slightly.   It transfers a  lower layer vector $g_t^l$ together with its future vectors $g_{t+\delta}^{l}$ into a new vector ${\zeta}_{t}^{l}$, where $\delta$ is future frame index. We modify the block of LSTM\_cuDNN or LSTM\_Custom with the context modeling. 
\begin{itemize}
	\item LSTM\_cuDNN\_Context:	the block is constructed with  a Nvidia cuDNN LSTM layer, followed by a linear projection layer, then the context modeling layer, and finally a LN layer.
	\item LSTM\_Custom\_Context: the block is constructed with the layer normalized LSTM layer with projection, and then followed by the context modeling layer.
\end{itemize}
A similar concept of context modeling was applied to RNN in \cite{wang2016lookahead} as Lookahead convolution layer. However, it was only applied to the top layer of a multi-layer RNN. In contrast, in this study we apply context modeling to every block of LSTM\_cuDNN or LSTM\_Custom, and also investigate its effectiveness in the context of E2E modeling. For RNN-T, we also investigate initializing the encoder with either CTC \cite{rao2017exploring} or CE training \cite{Hu2020}.  

RNN-AED models use blocks of LSTM\_cuDNN\_Context as encoder. Experiments with LSTM\_Custom\_Context will be a part of future study. The streaming mechanism we have chosen for this study is Monotonic Chunkwise Attention (MoChA) \cite{mocha}. MoChA consists of a monotonic attention mechanism \cite{monotonic_attention} which scans the encoder output in a left to right order and selects a particular encoder state when it decides to trigger the decoder. This selection probability is selected by sampling from a parameterized Bernoulli random variable. 
Once a trigger point is detected, MoChA also uses an additional lookback window and applies a regular softmax attention over that. Note that we have a sampling operation here, which precludes the use of standard backpropagation. Therefore we train with respect to the expected values of the context vectors. Please refer to \cite{mocha} for more details. 

To enable streaming scenario in Transformer-AED models, we borrow the idea in trigger-attention (TA) \cite{moritz2019triggered}, where the CTC conducts frame-synchronized decoding to select top-k candidates for each frame and then the  attention model is leveraged to jointly re-rank the candidates using  Eq. \ref{eq:joint_decoding} once a new subword is triggered by the CTC. Since the Transformer encoder is deeper than LSTM, the lookahead method may not be the best solution. We compare the chunk-based method and the lookahead-based method. The former segments the entire input into several fixed-length chunks and then feeds them into the model chunk by chunk, while the latter is exactly the same with the method in RNN-T and RNN-AED. For the chunk-based encoder, the decoder can see the end of a chunk. For the lookahead based encoder, we set a fixed window size for decoder.

\section{Experiments}
In this section, we evaluate the effectiveness of all models by  training them with 65 thousand (K) hours of transcribed Microsoft data. The test sets cover 13 application scenarios such as Cortana and far-field speech, containing a total of 1.8 million (M) words. We report the word error rate (WER) averaged over all test scenarios.  All the training and test data are anonymized with personally identifiable information removed. 

For fair comparison, all E2E models built for this study have around 87 M parameters. The input feature is 80-dimension log Mel filter bank with a stride of 10 milliseconds (ms). Three of them are stacked together to form a 240-dimension super-frame. This is fed to the encoder networks for RNN-T and RNN-AED, while Transformer-AED directly consumes the 10 ms feature. All E2E models use the same 4 K word piece units as the output target.

\subsection{Non-streaming E2E models}
As described in Section \ref{ssec:block}, the non-streaming RNN-T model uses bi-directional LSTM with Nvidia cuDNN library in its encoder. The LSTM memory cell size is 780. The LSTM outputs from the forward and backward direction are concatenated with the total dimension of 1560 then linearly projected to  dimension 780, followed by a LN layer. There are total 6 stacked blocks of such operation. The prediction network has 2 stacked blocks, each of which contains a uni-directional cuDNN LSTM with memory cell size of 1280, followed by a linear projection layer to reduce the dimension to 640, and then with a LN layer.  

The non-streaming RNN-AED model uses exactly the same encoder and decoder structures as the non-streaming RNN-T model. Similar to \cite{bahdanau2016end}, a location-aware attention mechanism is used. In addition to the encoder and decoder hidden states, this mechanism also takes alignments from previous decoder step as inputs. The attention dimension is 512.

The Transformer-AED model has 18 Transformer blocks in encoder and 6 Transformer blocks in decoder. Before Transformer blocks in encoder, we use a 4 layers VGG network to pre-process the speech feature with total stride 4.  The number of attention head is 8 and the attention dimension of each head is 64. The dimension of the feed-forward layer is 2048 in  Transformer blocks. The combination weights of joint training and decoding (i.e. $\alpha, \beta$) are both 0.3. 

As shown in Table \ref{tab:wer_non-streaming}, the non-streaming AED models have a clear advantage over the non-streaming RNN-T model due to the power of attention modeling. Transformer-AED improves RNN-AED by 2.7\% relative WER reduction. 

\begin{table}[t]
  \caption{Average WER of all non-streaming E2E models on 13 test sets containing 1.8 M words. }
  \label{tab:wer_non-streaming}
  \centering
  \begin{tabular}{l|c}
    	\hline
			non-streaming	models			& 	WER        \\							
    	\hline
		RNN-T (cuDNN)			& 9.25 \\
		RNN-AED (cuDNN)	&  8.05 \\ 
		Transformer-AED &  7.83 \\ \hline
    	\hline
  \end{tabular}
\end{table}

\subsection{Surpassing hybrid  model with streaming E2E models}
In \cite{li2020high} we reported results from our best hybrid model called the contextual layer trajectory LSTM (cltLSTM) \cite{li2019improving}. The cltLSTM was trained with a three-stage optimization process. This model was able to obtain a 16.2\% relative WER reduction over the CE baseline. Introducing 24 frames of total future-context further yields an 18.7\% relative WER reduction. The encode latency is only 480 ms (24*20ms=480 ms; stride-per-frame is 20 ms due to frame skipping \cite{Miao16}). Hence, this cltLSTM model (Table \ref{tab:wer_streaming}) presents a very challenging streaming hybrid model to beat. This model has 65 M parameters, and is decoded with 5 gigabytes 5gram decoding graph. 

We list the results for all streaming E2E models in Table \ref{tab:wer_streaming}. 
The baseline RNN-T implementation uses unidirectional cuDNN LSTMs in both the encoder and the decoder. The encoder has 6 stacked blocks of LSTM\_cuDNN. Each block has a unidirectional cuDNN LSTM with 1280 memory cells which projected to 640 dimension and followed by LN. The prediction and the joint network is the same as in the non-streaming RNN-T model.
This RNN-T model obtains 12.16\% test WER. The second RNN-T model inserts the context modeling layer (Eq. \eqref{eq:context}) after the linear projection layer in each block.  The context modeling has 4 frames lookahead at each block, and therefore the encoder has $4*6=24$ frames lookahead. Because the frame shift is 30 ms, the total encoder lookahead is 720ms. The lookahead brings great WER improvement, obtaining 10.65\% WER. This is 12.4\% relative WER reduction from the first RNN-T model without any lookahead. We also followed lookahead convolution proposed in \cite{wang2016lookahead} by using 24 frames lookahead only on the top most RNN block. This model gives 11.19\% WER, showing that our proposed context modeling, which allocates lookahead frames equally at each block, is better than lookahead convolution \cite{wang2016lookahead}, which simply puts all lookahead frames on the top layer only.   

Next, we look at the impact of encoder initialization for RNN-T. Shown in Table \ref{tab:wer_streaming}, the CTC initialization of RNN-T encoder doesn't help too much while the CE initialization significantly reduces WER to 9.80. This is 8.0\% relative WER reduction from the randomly initialized model. The CTC initialization makes the encoder  emit token spikes together with lots of blanks while CE initialization enables the encoder to learn time alignment. Given the gain with CE initialization,  we believe the encoder of RNN-T functions more like an acoustic model in the hybrid model. Note the CE pre-training needs time alignments, which is hard to get for word piece units as many of them don't have phoneme realisation. However, the time alignment for words is still accurate. We make an approximation and obtain alignments for a word piece by simply segmenting the duration of its word equally into its constituent word pieces.

For the last RNN-T model, we put projection layer and LN inside the LSTM cell (Custom\_LSTM), and then insert the context modeling layer after it. Putting projection layer inside allows us to use larger number of memory cells while keeping similar model size as the cuDNN\_LSTM setup. This LSTM has 2048 memory cells and the project layer reduces the output size to 640. 
This model finally gives 9.27\% WER, which is slightly better than our best hybrid model. 
   
\begin{table}[t]
  \caption{Average WERs of streaming models on 13 test sets containing 1.8 M words.}
  \label{tab:wer_streaming}
  \centering
  \begin{tabular}{l|c|c}
    	\hline
			streaming	models			& 	WER & encoder lookahead        \\							
    	\hline
    	hybrid & & \\
    	\quad cltLSTM & 9.34 & 480 ms \\
    	\hline \hline
		RNN-T  & & \\
		\quad cuDNN 			& 12.16 &  0 ms \\
		\quad cuDNN+Context 			& 10.65  & 720 ms \\
		\quad cuDNN+convolution \cite{wang2016lookahead} & 11.19 & 720 ms \\
		\quad cuDNN+Context+CTC init. 			& 10.62  & 720 ms \\
		\quad cuDNN+Context+CE init. 			& 9.80  & 720 ms \\
		\quad Custom+Context+CE init. 			& 9.27  & 720 ms \\		
		\hline
		RNN-AED &&\\
		\quad cuDNN+Context 	& 9.61 & 720 ms \\ 
		\hline
		Transformer-AED && \\
			\quad Lookahead method & 10.26 & 720 ms \\ 
			\quad Chunk-based method& 9.16 & 720 ms \\ 
    	\hline
  \end{tabular}
\end{table}

With the same encoder architecture as the cuDNN RNN-T, the MoChA-based streaming RNN-AED model gives impressive results. Unlike RNN-T, it does not need any initialization and is still able to slightly outperform it in an apple-to-apple comparison (9.61\% vs 9.80\%). To the best of our knowledge, this is the first time a streaming RNN-AED has outperformed RNN-T on a large scale task. Note that our previous study didn't observe  accuracy improvement for RNN-AED with CE initialization \cite{Hirofumi2020streaming}. We will investigate whether RNN-AED can also benefit from customized LSTM function in future study. 

The architecture of the streaming Transformer-AED model is the same as the non-streaming one.  For lookahead context-modeling method, each encoder block looks ahead  1 frame. Considering the total stride of VGG is 4 and the speech sampling rate is 10ms, the encoder has $1*18*4*10ms=720ms$ latency. The decoder of the lookahead method introduces an extra 240ms latency. The chunk-based method considers future context with a fixed-chunk. The latency of each frame is in the range of $[480\text{ms}, 960\text{ms}]$, resulting in a 720ms averaged latency without extra decoder latency. The chunk-based method obtains 9.16$\%$ WER, significantly outperforming the lookahead method, mainly because the bottom Transformer blocks of the lookahead approach cannot enjoy the full advantages provided by the right context. 



\section{Conclusions}
This work presents the first large-scale comparative study of three popular E2E models (RNN-T, RNN-AED, and Transformer-AED). The models are compared in both streaming and non-streaming modes. All models are trained with 65K hours of Microsoft's internal anonymized data. We observe that with the same encoder structure, AED is better than RNN-T for both non-streaming and streaming models. With customized LSTM and CE initialization for encoder, the RNN-T model becomes better than RNN-AED.  Among all models, Transformer-AED obtained the best WERs in both streaming and non-streaming modes. 

In this study, both streaming RNN-T and Transformer-AED outperformed a highly-optimized hybrid model. There are several significant factors contributing to this success. For streaming RNN-T, the proposed context modeling reduces the WER by 12.4\% relative from the one without any lookahead. The CE initialization for RNN-T improves over the random initialization baseline by 8.0\% relative WER reduction. This shows pretraining is helpful even on a large scale task. To utilize future context for streaming Transformer-AED, we show that the chunk-based method is better than the lookahead method by 10.7\% relative.  

\bibliographystyle{IEEEtran}

\bibliography{mybib}

\begin{thebibliography}{10}
\providecommand{\url}[1]{#1}
\csname url@samestyle\endcsname
\providecommand{\newblock}{\relax}
\providecommand{\bibinfo}[2]{#2}
\providecommand{\BIBentrySTDinterwordspacing}{\spaceskip=0pt\relax}
\providecommand{\BIBentryALTinterwordstretchfactor}{4}
\providecommand{\BIBentryALTinterwordspacing}{\spaceskip=\fontdimen2\font plus
\BIBentryALTinterwordstretchfactor\fontdimen3\font minus
  \fontdimen4\font\relax}
\providecommand{\BIBforeignlanguage}[2]{{%
\expandafter\ifx\csname l@#1\endcsname\relax
\typeout{** WARNING: IEEEtran.bst: No hyphenation pattern has been}%
\typeout{** loaded for the language `#1'. Using the pattern for}%
\typeout{** the default language instead.}%
\else
\language=\csname l@#1\endcsname
\fi
#2}}
\providecommand{\BIBdecl}{\relax}
\BIBdecl

\bibitem{DNN4ASR-hinton2012}
G.~Hinton, L.~Deng, D.~Yu \emph{et~al.}, ``Deep neural networks for acoustic
  modeling in speech recognition: The shared views of four research groups,''
  \emph{{IEEE} Signal Processing Magazine}, vol.~29, no.~6, pp. 82--97, 2012.

\bibitem{miao2015eesen}
Y.~Miao, M.~Gowayyed, and F.~Metze, ``{EESEN}: End-to-end speech recognition
  using deep {RNN} models and {WFST}-based decoding,'' in \emph{Proc.
  ASRU}.\hskip 1em plus 0.5em minus 0.4em\relax IEEE, 2015, pp. 167--174.

\bibitem{chan2016listen}
W.~Chan, N.~Jaitly, Q.~Le, and O.~Vinyals, ``Listen, attend and spell: A neural
  network for large vocabulary conversational speech recognition,'' in
  \emph{Proc. ICASSP}, 2016, pp. 4960--4964.

\bibitem{prabhavalkar2017comparison}
R.~Prabhavalkar, K.~Rao, T.~N. Sainath, B.~Li, L.~Johnson, and N.~Jaitly, ``A
  comparison of sequence-to-sequence models for speech recognition,'' in
  \emph{Proc. Interspeech}, 2017, pp. 939--943.

\bibitem{battenberg2017exploring}
E.~Battenberg, J.~Chen, R.~Child, A.~Coates, Y.~G.~Y. Li, H.~Liu, S.~Satheesh,
  A.~Sriram, and Z.~Zhu, ``Exploring neural transducers for end-to-end speech
  recognition,'' in \emph{Proc. ASRU}.\hskip 1em plus 0.5em minus 0.4em\relax
  IEEE, 2017, pp. 206--213.

\bibitem{rao2017exploring}
K.~Rao, H.~Sak, and R.~Prabhavalkar, ``Exploring architectures, data and units
  for streaming end-to-end speech recognition with {RNN}-transducer,'' in
  \emph{Proc. ASRU}, 2017.

\bibitem{chiu2018state}
C.-C. Chiu, T.~N. Sainath, Y.~Wu, R.~Prabhavalkar, P.~Nguyen, Z.~Chen,
  A.~Kannan, R.~J. Weiss, K.~Rao, K.~Gonina \emph{et~al.}, ``{State-of-the-art
  speech recognition with sequence-to-sequence models},'' in \emph{Proc.
  ICASSP}, 2018.

\bibitem{Li18CTCnoOOV}
J.~Li, G.~Ye, A.~Das, R.~Zhao, and Y.~Gong, ``Advancing acoustic-to-word {CTC}
  model,'' in \emph{Proc. ICASSP}, 2018.

\bibitem{he2019streaming}
Y.~He, T.~N. Sainath, R.~Prabhavalkar, I.~McGraw, R.~Alvarez, D.~Zhao,
  D.~Rybach, A.~Kannan, Y.~Wu, R.~Pang \emph{et~al.}, ``Streaming end-to-end
  speech recognition for mobile devices,'' in \emph{Proc. ICASSP}, 2019, pp.
  6381--6385.

\bibitem{Li2020Developing}
J.~Li, , R.~Zhao, Z.~Meng \emph{et~al.}, ``Developing {RNN-T} models surpassing
  high-performance hybrid models with customization capability,'' in
  \emph{Proc. Interspeech}, 2020.

\bibitem{graves2006connectionist}
A.~Graves, S.~Fern{\'a}ndez, F.~Gomez, and J.~Schmidhuber, ``Connectionist
  temporal classification: labelling unsegmented sequence data with recurrent
  neural networks,'' in \emph{Proceedings of ICML}.\hskip 1em plus 0.5em minus
  0.4em\relax ACM, 2006, pp. 369--376.

\bibitem{Graves-E2EASR}
A.~Graves and N.~Jaitley, ``Towards end-to-end speech recognition with
  recurrent neural networks,'' in \emph{PMLR}, 2014, pp. 1764--1772.

\bibitem{Graves-RNNSeqTransduction}
A.~Graves, ``Sequence transduction with recurrent neural networks,''
  \emph{CoRR}, vol. abs/1211.3711, 2012.

\bibitem{Attention-bahdanau2014}
D.~Bahdanau, K.~Cho, and Y.~Bengio, ``Neural machine translation by jointly
  learning to align and translate,'' \emph{arXiv preprint arXiv:1409.0473},
  2014.

\bibitem{Attention-speech-chorowski2015}
J.~K. Chorowski, D.~Bahdanau, D.~Serdyuk, K.~Cho, and Y.~Bengio,
  ``Attention-based models for speech recognition,'' in \emph{NIPS}, 2015, pp.
  577--585.

\bibitem{Sainath19}
T.~Sainath, R.~Pang, and et. al., ``Two-pass end-to-end speech recognition,''
  in \emph{Proc. Interspeech}, 2019.

\bibitem{Li2019RNNT}
J.~Li, R.~Zhao, H.~Hu, and Y.~Gong, ``Improving {RNN} transducer modeling for
  end-to-end speech recognition,'' in \emph{Proc. ASRU}, 2019.

\bibitem{jain2019rnn}
M.~Jain, K.~Schubert, J.~Mahadeokar \emph{et~al.}, ``{RNN-T} for latency
  controlled {ASR} with improved beam search,'' \emph{arXiv preprint
  arXiv:1911.01629}, 2019.

\bibitem{bahdanau2014neural}
D.~Bahdanau, K.~Cho, and Y.~Bengio, ``Neural machine translation by jointly
  learning to align and translate,'' \emph{arXiv preprint arXiv:1409.0473},
  2014.

\bibitem{chiu2017monotonic}
C.-C. Chiu and C.~Raffel, ``Monotonic chunkwise attention,'' \emph{arXiv
  preprint arXiv:1712.05382}, 2017.

\bibitem{moritz2019triggered}
N.~Moritz, T.~Hori, and J.~Le~Roux, ``Triggered attention for end-to-end speech
  recognition,'' in \emph{Proc. ICASSP}, 2019, pp. 5666--5670.

\bibitem{vaswani2017attention}
A.~Vaswani, N.~Shazeer, N.~Parmar, J.~Uszkoreit, L.~Jones, A.~N. Gomez,
  {\L}.~Kaiser, and I.~Polosukhin, ``Attention is all you need,'' in
  \emph{Advances in Neural Information Processing Systems}, 2017, pp.
  6000--6010.

\bibitem{dong2018speech}
L.~Dong, S.~Xu, and B.~Xu, ``Speech-transformer: a no-recurrence
  sequence-to-sequence model for speech recognition,'' in \emph{Proc. ICASSP},
  2018, pp. 5884--5888.

\bibitem{zhou2018syllable}
S.~Zhou, L.~Dong, S.~Xu, and B.~Xu, ``Syllable-based sequence-to-sequence
  speech recognition with the transformer in {Mandarin Chinese},'' in
  \emph{Proc. Interspeech}, 2018.

\bibitem{karita2019comparative}
S.~Karita, N.~Chen, T.~Hayashi \emph{et~al.}, ``A comparative study on
  transformer vs {RNN} in speech applications,'' in \emph{Proc. ASRU}, 2019.

\bibitem{chiu2019comparison}
C.-C. Chiu, W.~Han, Y.~Zhang \emph{et~al.}, ``A comparison of end-to-end models
  for long-form speech recognition,'' in \emph{Proc. ASRU}, 2019.

\bibitem{sainath2020streaming}
T.~N. Sainath, Y.~He, B.~Li \emph{et~al.}, ``A streaming on-device end-to-end
  model surpassing server-side conventional model quality and latency,'' in
  \emph{Proc. ICASSP}, 2020, pp. 6059--6063.

\bibitem{li2020high}
J.~Li, R.~Zhao, E.~Sun, J.~H. Wong, A.~Das, Z.~Meng, and Y.~Gong,
  ``High-accuracy and low-latency speech recognition with two-head contextual
  layer trajectory {LSTM} model,'' in \emph{Proc. ICASSP}, 2020.

\bibitem{li2019improving}
J.~Li, L.~Lu, C.~Liu, and Y.~Gong, ``Improving layer trajectory {LSTM} with
  future context frames,'' in \emph{Proc. ICASSP}, 2019, pp. 6550--6554.

\bibitem{Wang2020Transformer}
\BIBentryALTinterwordspacing
C.~Wang, \emph{Streaming Transformer}, 2020. [Online]. Available:
  \url{https://github.com/cywang97/StreamingTransformer}
\BIBentrySTDinterwordspacing

\bibitem{Hochreiter1997long}
S.~Hochreiter and J.~Schmidhuber, ``Long short-term memory,'' \emph{Neural
  computation}, vol.~9, no.~8, pp. 1735--1780, 1997.

\bibitem{zhang2020transformer}
Q.~Zhang, H.~Lu, H.~Sak \emph{et~al.}, ``Transformer transducer: A streamable
  speech recognition model with transformer encoders and {RNN-T} loss,'' in
  \emph{Proc. ICASSP}, 2020.

\bibitem{gulati2020conformer}
A.~Gulati, J.~Qin, C.-C. Chiu,  \emph{et~al.}, ``Conformer:
  Convolution-augmented transformer for speech recognition,'' \emph{arXiv
  preprint arXiv:2005.08100}, 2020.

\bibitem{bahdanau2016end}
D.~Bahdanau, J.~Chorowski, D.~Serdyuk, P.~Brakel, and Y.~Bengio, ``End-to-end
  attention-based large vocabulary speech recognition,'' in \emph{Proc.
  ICASSP}.\hskip 1em plus 0.5em minus 0.4em\relax IEEE, 2016, pp. 4945--4949.

\bibitem{RESNET-he2015}
K.~He, X.~Zhang, S.~Ren, and J.~Sun, ``Deep residual learning for image
  recognition,'' \emph{arXiv preprint arXiv:1512.03385}, 2015.

\bibitem{ba2016layer}
J.~L. Ba, J.~R. Kiros, and G.~E. Hinton, ``Layer normalization,'' \emph{arXiv
  preprint arXiv:1607.06450}, 2016.

\bibitem{chetlur2014cudnn}
S.~Chetlur, C.~Woolley, P.~Vandermersch, J.~Cohen, J.~Tran, B.~Catanzaro, and
  E.~Shelhamer, ``{cuDNN}: Efficient primitives for deep learning,''
  \emph{arXiv preprint arXiv:1410.0759}, 2014.

\bibitem{wang2020semantic}
C.~Wang, Y.~Wu, Y.~Du, J.~Li, S.~Liu, L.~Lu, S.~Ren, G.~Ye, S.~Zhao, and
  M.~Zhou, ``Semantic mask for transformer based end-to-end speech
  recognition,'' in \emph{Proc. Interspeech}, 2020.

\bibitem{simonyan2014very}
K.~Simonyan and A.~Zisserman, ``Very deep convolutional networks for
  large-scale image recognition,'' in \emph{Proc. ICLR}, 2015.

\bibitem{wang2016lookahead}
C.~Wang, D.~Yogatama, A.~Coates, T.~Han, A.~Hannun, and B.~Xiao, ``Lookahead
  convolution layer for unidirectional recurrent neural networks,'' in
  \emph{Proc. ICLR Workshop}, 2016.

\bibitem{Hu2020}
H.~Hu, R.~Zhao, J.~Li, L.~Lu, and Y.~Gong, ``Exploring pre-training with
  alignments for {RNN} transducer based end-to-end speech recognition,'' in
  \emph{Proc. ICASSP}, 2020.

\bibitem{mocha}
C.-C. Chiu* and C.~Raffel*, ``Monotonic chunkwise attention,'' in
  \emph{International Conference on Learning Representations}, 2018.

\bibitem{monotonic_attention}
C.~Raffel, D.~Eck, P.~Liu, R.~J. Weiss, and T.~Luong, ``Online and linear-time
  attention by enforcing monotonic alignments,'' in \emph{Thirty-fourth
  International Conference on Machine Learning}, 2017.

\bibitem{Miao16}
Y.~Miao, J.~Li, Y.~Wang, S.~Zhang, and Y.~Gong, ``Simplifying long short-term
  memory acoustic models for fast training and decoding,'' in \emph{Proc.
  ICASSP}, 2016.

\bibitem{Hirofumi2020streaming}
H.~Inaguma, Y.~Gaur, L.~Lu, J.~Li, and Y.~Gong, ``Minimum latency training
  strategies for streaming sequence-to-sequence asr,'' in \emph{Proc. ICASSP},
  2020.

\end{thebibliography}

\end{document}